
\input phyzzx 
%
%
\newcount\lemnumber   \lemnumber=0
\newcount\thnumber   \thnumber=0
\newcount\conumber   \conumber=0

\def\myeq{{\the\equanumber}}

\def\Lemma{\par\noindent\global\advance\lemnumber by 1
           {\bf Lemma\ (\chapterlabel\the\lemnumber)}}
\def\Corollary{\par\noindent\global\advance\conumber by 1
           {\bf Corollary\ (\chapterlabel\the\conumber)}}
\def\Theorem{\par\noindent\global\advance\thnumber by 1
           {\bf Theorem\ (\chapterlabel\the\thnumber)}}

%
%
\def\e{\adveq\eqno{\rm (\chapterlabel\the\equanumber)}}
\def\adveq{\global\advance\equanumber by 1}
\def\twoline#1#2{\displaylines{\qquad#1\hfill(\adveq\myeq)\cr\hfill#2
\qquad\cr}}


%
%
\font\tensl=cmsl10
\font\tenss=cmssq8 scaled\magstep1
\outer\def\quote{
   \begingroup\bigskip\vfill
   \def\endquote{\endgroup\eject}
    \def\par{\ifhmode\/\endgraf\fi}\obeylines
    \tenrm \let\tt=\twelvett
    \baselineskip=10pt \interlinepenalty=1000
    \leftskip=0pt plus 60pc minus \parindent \parfillskip=0pt
     \let\rm=\tenss \let\sl=\tensl \everypar{\sl}}
\def\from#1(#2){\smallskip\noindent\rm--- #1\unskip\enspace(#2)\bigskip}

\def\CALT{\address{Division of Physics, Mathematics
and Astronomy\break
Mail Code 452--48\break
California Institute of Technology\break
Pasadena, CA 91125}}

\def\r#1{$\lb \rm#1 \rb$}

%
%
\def\rarrow{\rightarrow}

\def\semidirect{\mathrel{\raise0.04cm\hbox{${\scriptscriptstyle |\!}$
\hskip-0.175cm}\times}}

\def\mod{\mathop{\rm mod}\nolimits}

\def\ref#1{$^{#1}$}

\def\pr#1{{#1^\prime}}

\def\half{{1\over2}}
\def\lb{\lbrack}
\def\rb{\rbrack}

\def\pr{\prime}
\def\diam{{\hbox{\hskip-0.02in
\raise-0.126in\hbox{$\displaystyle\bigvee$}\hskip-0.241in
\raise0.099in\hbox{ $\displaystyle{\bigwedge}$}}}}

\def\bw#1#2#3#4#5{{w\left(\matrix{#1&#2\cr#3&#4\cr}\bigg\vert #5\right)}}

\def\prr#1{{#1^{\prime\prime}}}
\def\pr#1{{#1^\prime}}
\def\ident{\equiv}
\overfullrule=0pt
\date{June, 1993}
\titlepage
\title{Graph IRF Models and Fusion Rings}
\author{Doron Gepner\foot{On leave of absence from the Weizmann Institute.
Incumbent of the Soretta and Henry Shapiro Chair.}}
\CALT
\vskip15pt
\abstract
Recently, a class of interaction round the face (IRF) solvable lattice
models were introduced, based on any rational conformal field theory
(RCFT). We investigate here the connection between the general
solvable IRF models and the fusion ones. To this end, we introduce
an associative algebra associated to any graph, as the algebra of
products of the eigenvalues of the incidence matrix. If a model is
based on an RCFT, its associated graph algebra is the fusion ring of the
RCFT. A number of examples are studied. The Gordon--generalized IRF
models are studied, and are shown to come from RCFT, by the graph algebra
construction. The IRF models based on the Dynkin diagrams of A-D-E are
studied. While the $A$ case stems from an RCFT, it is shown that
the $D-E$ cases do not. The graph algebras are constructed, and
it is speculated that a natural isomorphism relating these to RCFT
exists. The question whether all solvable IRF models stems from an RCFT
remains open, though the $D-E$ cases shows that a mixing of the
primary fields is needed.
\endpage
\par
Recently, this author has put forward four categorical isomorphisms among
\REF\Found{D. Gepner, ``Foundations of rational conformal field theory, I'',
Caltech preprint, CALT--68--1825, November (1992).}
important problems that arise in two dimensional physics \r\Found.
These are
integrable $N=2$ supersymmetric models, rational conformal field theories
(RCFT), fusion interaction round the face models (IRF), and integrable
soliton systems. It has been shown that the latter three categories are
equivalent, and evidence was given for the equivalence
with the first category.

The purpose of this note is to further explore these equivalences. In
particular, many solvable IRF models are known to be connected
to certain (very special) graphs. The graphs are the admissibility
conditions for the state variables that are allowed to be on the same
link on the lattice. One could try to enlarge the isomorphisms
mentioned above to {\bf all} solvable IRF lattice models.
This would mean that the graph must be obtained from the fusion ring
of some rational conformal field theory.
Specifically,
this raises the question: is an IRF model solvable if and only
if its admissibility graph arises from the fusion ring of some
RCFT? In this note we wish investigate this question.
In particular, for each graph we will
associate a commutative algebra, which is essentially the multiplication
table for the eigenvalues of the incidence matrix. If our conjecture holds,
this very same ring must be the fusion ring of some rational conformal
field theory. We shall explore a variety of examples.

Let the pair $(S,K)$ where $K\in S\times S$ be a graph, where $S$ is
the set of points of the graph and $K$ is the incidence matrix. More
generally, we shall allow oriented graphs with multiple links, more
conveniently
described by the incidence matrix $M_{i,j}$, whose non--negative integer
entries describe the number of links from the point $i$ to $j$ where
$i,j\in S$. For such a graph we can associate (ambiguously)
an interaction round the face (IRF) lattice model. The partition function
of the model, which is defined on a square lattice, is given by
\def\bwnou#1#2#3#4{w\pmatrix{#1&#2\cr #3& #4\cr}}
$$Z=\sum_{\rm states} \prod_{\rm faces} \bwnou a b c d,\e$$
where $a,b,c,d$ are the state variables, and $a$ and $b$ are allowed to be
on the same link iff $a$ and $b$ are connected by the graph, denoted by
$a\sim b$. $\bwnou a b c d$ is some Boltzmann weight, which is still undefined,
and can be considered as the parameters of the model. For multiple links,
the Boltzmann weight depends on the particular link, in an obvious
generalization. For some graphs and some choices of Boltzmann weights the
models are solvable, in the sense that two other Boltzmann weights
$\pr w$ and $\prr w$ can be found, obeying the same admissibility condition,
such that the following relation holds,
$$\twoline{
\sum_c w\pmatrix{b&d\cr a&c\cr}\pr w\pmatrix{a&c\cr g&f\cr} \prr w \pmatrix{
c&c\cr f&e\cr}=}{
\sum_c \prr w\pmatrix{a&b\cr g&c\cr}\pr w\pmatrix{b&d\cr c&e\cr}
w\pmatrix{c&e \cr g&f\cr}}$$
This relation is called the star--triangle equation (STE). It is a
very powerful
tool in the calculation of the partition function eq. (1), and forms the
basis for the solvability of the model.

This raises the important question of which graphs and which choices
of Boltzmann
weights lead to solvable IRF models. In fact, study have shown, that only
for very special graphs such solvable Boltzmann weights exists at all,
and then they are more or less unique
\REF\Baxter{R.J. Baxter, ``Exactly solved models in statistical mechanics'',
Academic Press, London, 1982.} (for a review, see e.g., \r\Baxter).
One might speculate that such a solution
exists if and only if the graph in question corresponds to the fusion
ring of some RCFT, and then the Boltzmann weights are described uniquely
by the braiding matrices of the RCFT. Actually, there are obvious counter
examples
to this conjecture. However, these IRF models do not have second order
phase transition points and thus can be considered as `bad' models
in the aforementioned sense. Precisely put: does all solvable IRF models
with a second order fixed point stem from an RCFT, in the above sense?

Let us thus delve into the definition of fusion IRF models.
Let $\cal O$ be a rational conformal field theory, and let $x$ be
a field, typically primary, in the theory. For an explanation of these
notions see for example \r\Found. In such a theory, the fusion of
the primary fields defines a commutative semi--simple ring,
$$[p] \times [q]=\sum_r N^r_{p,q} [r],\e$$
where $[p],[q],[r]$ denote the primary fields, and $N_{p,q}^r$ are
the structure constants, which are non--negative integers.
Now, for any such ring
we can associate a family of graphs in the following fashion.
We let the points of the graph be the primary fields of the
theory, and we identify the incidence matrix $M_{p,q}$ with
the structure constants with respect to a fixed field in the theory $[x]$,
$M_{p,q}=N_{x,p}^q$. Now, given such a pair $({\cal O},x)$, we can
define an IRF model based of the fusion graph of the field $x$, i.e.,
$p\sim q$ iff $N_{xp}^q>0$. We denote the resulting lattice model by
IRF$({\cal O},x)$.
It was shown in ref. \r\Found\ that indeed all such
models, termed fusion IRF models are solvable, and that Boltzmann
weights satisfying the STE, eq. (2), can be found. The Boltzmann
weights are extensions of the braiding matrices of the corresponding RCFT.
The questions is then, is the converse true and all such solvable models
are fusion IRF?

In any event, we can examine known solvable IRF models, to determine if
their admissibility graph comes from an RCFT. If this is the case, such graphs
has to obey some very special properties that are nearly enough to settle
the question, case by case, as well as to determine the specific RCFT.

\REF\Ver{E. Verlinde, Nucl. Phys. B300 (1988) 360}
It was shown in ref. \r\Ver\ that the fusion ring in an RCFT is connected
to a unitary matrix which is the matrix of modular transformations.
The important thing about $S$ is that it diagonalizes the eigenvalues
of the fusion ring. Namely, if we define
$$\{ i\}=\sum_{j} {S_{i,j}\over S_{i,0}} [j],\e$$
then $\{ i\}$ obeys the fusion product
$$\{ i\} \times \{j \}=\delta_{ij} \{i\}.\e$$
As the matrix $S$ is unitary, this determines it uniquely from the fusion ring,
up to a permutation of the rows, as it is simply the matrix that diagonalizes
the fusion ring. (More precisely, it is the point basis in the affine variety
defined by the ring
\REF\GepCom{D. Gepner, Comm. Math. Phys. 141 (1991) 381.}
\r\GepCom). However, not all rings lead to sensible
$S$ matrices, and those that do are very special. The reason is that
in RCFT, the $S$ matrix needs to be symmetric, and not just unitary,
$S=S^t$. This alone is a very strong constraint on the allowed
fusion rings. A further restriction arises from the fact that every
such ring must admit a non--degenerate
symmetric bi--linear form $(a,b)$, where
$a$ and $b$ are primary fields, defined by $(a,b)=1$ iff $N_{ab}^1=(a,b)$,
where
$1$ stands for the unit in the ring (which is a primary field). Further
$(a,b)$ must be either $0$ or $1$ for all the primary fields $a$ and $b$,
and for each primary $a$, $(a,b)$ is zero, for all $b$ except for a unique
choice. Thus the bi-linear form defines a unique conjugate for each field,
$\bar a$ which is the unique field for which $(a,\bar a)=1$.

Thus, the question whether a given graph stems from an RCFT can be examined
on the basis of whether the above properties holds for the graph, and its
associated fusion ring. Let $(S,K)$ be an arbitrary graph then, with the
incidence matrix $M_{i,j}$. Denote the eigenvalues of $M$ by $v_j^\alpha$,
i.e.,
$$\sum_j M_{i,j} v_j^\alpha=\sum_\gamma \lambda^\alpha v_i.\e$$
We can normalize the eigenvalues to unity, $\sum_j v_j^\alpha
{v_j^\alpha}^\dagger=1$. The eigenvalues are thus uniquely defined
(up to a phase). We can now write down a commutative associative algebra
associated to the eigenvalues. We do so by specifying a unique choice for
the `unit' element, denoted by say $1$. Further, we define the product of the
elements
$\alpha$ and $\beta$ to be,
$$[\alpha]\times [\beta]=\sum_\gamma N_{\alpha,\beta}^\gamma [\gamma],\e$$
where the structure constants $N_{\alpha,\beta}^\gamma$ are defined by
$${v_j^\alpha\over v_j^1} {v_j^\beta\over v_j^1}=
\sum_\gamma N_{\alpha,\beta}^\gamma {v^\gamma_j\over v_j^1},\e$$
for all $j$. Since the eigenvectors are linearly independent, $N$ is
so uniquely defined. The eigenvalues can be normalized, in which case the
matrix $v_j^h$ is unitary,
$$\sum_j v_j^h (v_j^p)^*=\delta_{h,p},\e$$
where $h$ and $p$ are any two exponents. We thus find from eq. (8), the
following form for the structure constants,
$$N_{p,q}^r=\sum_j {v^p_j v^q_j (v^r_j)^* \over v^1_j}.\e$$
If $v_j^h$ is a modular matrix of an RCFT
\REF\GW{D. Gepner and E. Witten, Nucl. Phys. B278 (1986) 493.}
\r\GW, eq. (10) gives
the fusion coefficients \r\GW, according to the formula of ref. \r\Ver.

For a non RCFT, since the matrix of eigenvalues, $v_j^p$, is
inherently non--symmetric, we can define a transposed algebra,
based on the nodes of the diagram, instead, in a similar fashion.
The structure constants of the algebra are then given by
$$M_{ij}^k=\sum_h {v_i^h v_j^h (v_k^h)^* \over v^h_1},\e$$
where the structure constants, $M_{ij}^k$ describe the product
of the nodes of the graph,
$${v_i^h\over v_1^h} {v_j^h\over v_1^h}=\sum_k M_{ij}^k {v_k^h\over
v_k^1}.\e$$
For an RCFT the two algebras are, of course, the same.
The algebra so defined suffers from a number
of ambiguities. First, the phase of the eigenvectors was arbitrary.
However, this is simply a redefinition of the basis elements. More
importantly the choice for the unit field `$1$' was arbitrary, and for
each such choice a different algebra is found. To summarize, for a
pair of any graph and a point in it we defined uniquely an algebra,
denoted by $A(G,p)$, where $G$ is the graph and $p$ is the point.
Further, the algebra has a unique preferred basis, up to a phase.

Now, if the graph in question stems from a fusion ring,
then the graph algebra, so defined, is identical with the fusion ring of the
theory, provided that we take for the preferred point the unit field of the
fusion ring. Further, up to a phase, the preferred basis of the graph algebra
is the primary field basis of the fusion ring, up to the phase ambiguity
mentioned above.

It follows that the question whether an IRF model stems from an RCFT
boils down to the question of whether its associated graph algebra
is a fusion ring. In light, of the many properties of such fusion rings,
only very special graphs can be candidates for fusion rings. Further,
the RCFT may be constructed from the fusion ring itself. Thus by
studying the graph algebra, the question raised in the introduction
can be settled.

Let us illustrate this construction by an example. A class of solvable
IRF models called the Gordon--Generalized (GG) hierarchy has been found
\REF\GG{A. Kuniba, Y. Akutsu and M. Wadati, J. Phys. Soc. Jpn 55 (1986)
 1092, Phys. Lett. A 116 (1986) 382, Phys. Lett. A 117 (1986) 358;
R.J. Baxter and G.E. Andrews, J. Stat. Phys. 44(1986) 249, 371.}
\r\GG.
The state variables in these models take the values $a=0,1,\ldots,k-1$, where
$a$ are the state variables, and $k$ is any integer. The admissibility
condition for the graph is
$$a\sim b \qquad {\rm iff}\ a+b\leq k-1.\e$$
It was shown in ref. \r\GG\ that the models so obtained are solvable,
and Boltzmann weights satisfying the STE were found. The case of $k=2$
is the well known hard hexagon model
\REF\HardHex{R.J. Baxter, J. Phys. A 13 (1980) L61.}
\r\HardHex. Now, let us construct
the graph algebra associated to this graph, for any $k$. For the unit field
we choose the element $[k-1]$. It is a straight forward calculation
that the algebra so obtained assumes the form,
$$[i]\times [j]=\sum_{m=|i-j|\atop m-i-j=0\mod2}^{2k-1-i-j} [m],\e$$
where we identified $[2k-1-i]\ident [i]$. It can be checked that this
algebra has all the properties of a fusion ring. In fact, this is
the known fusion ring of the RCFT $SU(2)_{2k-1}/SU(2)_{1/(2k-1)}$,
described in ref. \r\Found. The graph itself is obtained from the field
$x=[k-1]$. We conclude that the GG hierarchy is the fusion IRF model
IRF$(SU(2)_{2k-1}/SU(2)_{1/(2k-1)},[k-1])$. It can be further checked that
the Boltzmann weights described in ref. \r\GG\ are indeed the extensions
of the braiding matrices of this RCFT.

It is quite straight forward to see directly that the admissibility
condition for the GG hierarchy models, eq. (13), is indeed precisely what is
obtained by fusion with respect to the field $[k-1]$ in the theory
$G=SU(2)_{2k-1}/SU(2)_{1/(2k-1)}$. We identify the state $(\sigma)$
of the GG hierarchy model with the primary field $[k-1-\sigma]$ in the
theory $G$. We can now compute the fusion with respect to the field
$[k-1]$. We find, according to eq. (14),
$$[k-1]\times [k-1-\sigma]=[\sigma]+[\sigma+2]+\ldots +[2k-2-\sigma]=
\sum_{\rho=0}^{k-1-\rho} [k-1-\rho],\e$$
where we used the identification of fields, $[\sigma]=[2k-1-\sigma]$, which
holds in the theory $G$.
As the state $(\sigma)$ is identified with the primary field $[k-1-\sigma]$,
eq. (15) implies precisely the GG admissibility condition, eq. (13).
This concludes the proof that the GG models are fusion IRF.

The theory $G$ may be constructed explicitly, ref. \r\Found, as
a sub--sector of the theory $SU(2)_{2k-1}\times (E_7)_1$, with an
extended algebra (for an example). The case of $k=1$ corresponds to
$(G_2)_1$ current algebra.

Consider now, as another example, the KAW hierarchy of models defined
in ref.
\REF\KAW{A. Kuniba, Y. Akutsu and M. Wadati, J. Phys. Soc. Jpn. 55 (1986)
2605.} \r\KAW. The state variables of the models along with their admissibility
condition are given by,
$$\sigma_i=0,1,2,\ldots,k-1,\qquad k-2\leq \sigma_i+\sigma_j\leq k,\e$$
where $k$ is some integer, and $\sigma_i$ and $\sigma_j$ are any two
neighboring states. Let $G\equiv SU(2)_k$ be the current algebra theory
associated to $SU(2)_k$. Denote as before by $[j]$ the field with the
isospin $j/2$, and let $p=[k-1]+[k]$ be a field which is a mixture of
two primary fields. We can compute the fusion with respect to the
field $p$, according to the usual rules of $SU(2)_k$, eq. (24), and we find,
$$p\times [\sigma_i]=[\sigma_i]\times [k]+\sigma_i\times [k-1]=
\sum_{\sigma_j \atop k-2\leq \sigma_i+\sigma_j\leq k} [\sigma_j],\e$$
which are precisely the fusion admissibility conditions of the KAW
hierarchy, eq. (16). Thus, we conclude that the GAW model is the
fusion lattice model IRF$(SU(2)_k,[k]+[k-1],[k]+[k-1])$. This is
an example of a model based on a mixture of primary fields. Such models
where considered in ref. \r\Found, and their Boltzmann weights given as
an extension of the conformal braiding matrices of the RCFT. It would
be an interesting exercise to compare the Boltzmann weights given
in ref. \r\Found, based on the fusion properties, and those given in
ref. \r\KAW, by a direct solution, and to show that they indeed coincide.

Let us turn now to another example. Consider the so called grand hierarchy
of solvable IRF lattice models, discussed in refs.
\REF\Grand{Y. Akutsu, A. Kuniba and M. Wadati, J. Phys. Soc. Jpn. 55 (1986)
2907; E. Date, M. Jimbo, T. Miwa and M. Okado, Lett. Math. Phys. 12 (1986) 209;
Phys. Rev. B35 (1987) 2105; E. Date, M. Jimbo, A. Kuniba, T. Miwa and
M. Okado, Nucl. Phys. B290 [FS20] (1987) 231}\r{\KAW,\Grand}.
These models are described by
$$l_i=0,1,\ldots, k,\qquad l_i-l_j=-N,-N+2,\ldots, N,\qquad l_i+l_j=N,
N+2,\ldots, k-N,\e$$
where $l_i$ are the state variables, $l_i$ and $l_j$ are adjacent,
and $k$ and $N$ are arbitrary
integers. For each $k$ and $N$, a solvable model was found \r\Grand, using
compositions of the eight vertex model. In fact, eq. (18), is exactly
the well known fusion rules of $SU(2)_k$.
It is rather evident from these fusion rules, eq. (24), that $l_i$
is admissible to $l_j$, if and only if $N_{l_j,p}^{l_i}\geq0$,
where $p=[N]$ primary field, in the previous notation. We conclude that
the grand hierarchy is exactly the fusion IRF models IRF$(SU(2)_k,[N],[N])$,
for any $k$ and any $N$. Again, it would be interesting to
verify that the Boltzmann weights coming from RCFT \r\Found, and those
computed directly in ref. \r\Grand, are identical. For $N=1$ (the
\REF\ABF{G.E. Andrews, R.J. Baxter and P.J. Forrester, J.
Stat. Phys. 35 (1984) 193}
Andrews--Baxter--Forrester model \r\ABF), this was done in
ref. \r\Found, and the results indeed agree. For larger $N$ this verification
is left to further work.

Before proceeding, let us discuss one subtlety in the logic of identifying the
graph algebra with the fusion rules. Recall that the primary fields were
identified with the eigenvalues of the incidence matrix, and that this
identification was unambiguous up to a phase. However, in case the incidence
matrix has degenerate eigenvalues, we can no longer distinguish which
mixture of these eigenvectors are the primary fields, and additional
information may be needed.

Let us now proceed to another interesting family of solvable IRF lattice
models. These are the lattice models based on the simple Lie algebras
which are $A_n$
\r\ABF\ and $D_n$, $E_6$, $E_7$ and $E_8$
\REF\Pas{V. Pasquier, J. Phys. A20 (1986) L217, L221}\r\Pas.
The states of the ADE IRF models
are in one--to--one correspondence with the simple roots of the respective
Lie algebra. Similarly, the admissibility
graph is the Dynkin diagram of the algebra. Thus, the incidence matrix of
the model is given by $M_{ab}=2\delta_{ab}-C_{ab}$ where $C_{ab}$ is
the Cartan matrix of the algebra. (For a review on simple Lie algebras
see, e.g., \REF\Hump{Humphreys, Introduction to Lie algebras and representation
theory, Springer--Verlag, New--York (1972)}
\r\Hump.)

The Boltzmann weights of the ADE models \r\Pas\ have the relatively simple
graph
state form (see, e.g., \r\Found\ and ref. therein),
$$\bw a b c d u=\sin(\lambda-u)\delta_{b c}+\left({\psi_b\psi_c\over \psi_a
\psi_d}\right) ^\half \sin u,\e$$
where $u$ labels the different Boltzmann weights satisfying the STE, eq. (2),
which are given by the values $u$, $u+v$ and $v$, for $w$, $\pr w$ and
$\prr w$, respectively, for any complex $u$ and $v$. $\psi_a$ is the
eigenvector
with the largest eigenvalue of the incidence matrix (the so called
Perron--Frobenius vector),
$$\sum_b M_{ab} \psi_b= 2\cos\lambda \psi_a,\e$$
where $\beta=2\cos\lambda$ is the maximal eigenvalue of the incidence matrix,
given by
$$\lambda={\pi\over g},\e$$
where $g$ is the Coxeter number of the algebra.
The entire set of eigenvalues of the incidence matrix is given by
$$\lambda_h=2\cos(\pi h/g),\e$$
where $h$ is any of the exponents of the Lie algebra (which can be degenerate).
The exponents, along with the Coxeter number are described in table (1).

%
%
\topinsert
\line{\hfill Table 1.\hfill}
\vskip15pt
\line{\hfill
\vbox{\offinterlineskip\hrule
\halign{\vrule\strut\quad \hfil#\quad\hfil&\vrule\quad\hfil # \hfil
\quad&\vrule\quad #\quad&\vrule#\cr
\noalign{\hrule}
Algebra&Coxeter Number&Exponents&\cr
\noalign{\hrule}
$A_n$&$n+1$&$1,2,\ldots,n$&\cr
$D_n$&$2n-2$&$1,3,5,\ldots,2n-3,n-1$&\cr
$E_6$&$12$&$1,4,5,7,8,11$&\cr
$E_7$&$18$&$1,5,7,9,11,13,17$&\cr
$E_8$&$30$&$1,7,11,13,17,19,23,29$&\cr
}\hrule}
\hfill}
\vskip20pt\endinsert

Now, in light of the conjecture raised in the introduction, we
would like
to examine if the ADE IRF models are fusion IRF models, i.e. if
they arise from a conformal field theory. To this end, let us
proceed with constructing the graph algebras associated with the
Dynkin diagrams of simple Lie algebras. If our conjecture is correct,
this should be the fusion ring of some RCFT. To do so, we first need
to calculate the eigenvectors of the Cartan matrix of each Lie algebra,
and then insert these into eq. (10).

We shall skip the $A_n$ cases (ABF models), as these
have already been demonstrated to be the fusion IRF models associated
with the RCFT $SU(2)_{n-1}$ \r\Found.
For completeness sake, the eigenvectors
for the $A_n$ graph are given by
$$v_{ij}=\left( {2\over n+1}\right)^\half  \sin({\pi ij\over
n+1}),\e$$
where $i$ labels the simple roots and $j$ labels the exponents, and
$i,j=1,2,\ldots,n$. This is non--else but the toroidal modular matrix
of the RCFT $SU(2)_k$ \r\GW, showing that the graph algebra is identical
to the fusion ring of the model, which has the product rule \r\GW,
$$[i]\times [j]=\sum_{l=|i-j| \atop l-i-j=0\mod 2}^{\min(2k-i-j,i+j)} [l],\e$$
where $[l]$ labels the $l$th primary field. The Boltzmann weight, eq. (10),
can be seen to give at the limit $u\rarrow i\infty$ the
braiding matrix of the respective RCFT \r\Found, concluding the proof
that the ABF model is a fusion IRF.

Let us turn now to the $D_n$ algebras. From table (1) the exponents
of the algebra are,
$1,3,5,\ldots,2n-3,n-1$, and thus are all different for odd $n$,
and have a twofold degeneracy at $n-1$ for even $n$. The eigenvalues of
the incidence matrix are given by eq. (22), $\lambda_h=2\cos(\pi h/(2n-2))$,
where $h$ is any of the exponents. The eigenvectors of the incidence matrix
are readily computed and are found to be,
$$v_j^h=\cases{\sqrt{2\over n-1} \sin({\pi h j\over 2n-2}) & {\rm for\ }
               $j\leq n-2$\cr
               {(-1)^j\over\sqrt{2(n-1)}}& {\rm for\ }$j=n-1,n-2$,\cr }\e$$
for the odd exponents $h=1,3,5,\ldots,2n-3$. For the exceptional exponent,
$h=n-1$, we find,
$$v_j^{n-1}=\cases{0& {\rm for \ } $j\leq n-2$,\cr
                   (-1)^j\over \sqrt 2 &{\rm for\ }$j\geq n-1$.\cr}\e$$
We have normalized the eigenvectors to have the absolute value one,
and thus $v_j^h$ is the unitary matrix which diagonalizes the incidence
matrix.

We next proceed to calculate the graph algebra, using eq. (10). Denote
by $[h]$, $h=1,3,\ldots 2n-3$, the elements of the algebra associated to
the regular exponents, and by $z=[n-1]$ the element associated to the
exceptional one.
Then the graph algebra can be computed from eq. (10), and we find,
$$\eqalign{
[p]\times [q]&=\sum_{r=|p-q|+1 \atop r=1\mod 2}^{\min(p+q-1,4n-5-p-q)}
[r],\cr
z\times [p]&=(-1)^{(p-1)/2} z,\cr
z\times z&={1\over 2(n-1)}\sum_{r=1\atop r=1\mod 2}^{2n-3} (-1)^{(r-1)/2}
[r].\cr}\e$$
It is striking that up to a trivial rescaling of $z$, $z\rarrow
z\sqrt{2(n-1)}$, all the structure constants are integers.
This implies that the graph algebra of $D_n$, any $n$, is actually
a commutative ring with a unit. As no two eigenvectors are the same,
the ring is a semi--simple one, which is a finite dimensional algebra,
with vanishing nil and Jacobson radicals. It is straight forwards to present
this ring in terms of generators and relations. Let $T_n(x)$ be the
Chebishev polynomial of the second kind, defined by $T_n(2\cos \phi)=
{\sin[(n+1)\phi]\over \sin \phi}$. Then the generators of the ring may be taken
to be $z$ and $x=[3]$, along with the relations $p_1(x)=p_2(z,x)=p_3(z,x)=0$,
where
$$\eqalign{p_1(x)&=T_{2n-4}(\sqrt{1+x})+T_{2n-2}(\sqrt{1+x}),\cr
           p_2(x,z)&=z^2-{1\over2(n-1)}\sum_{h=1\atop h=1\mod 2}^{2n-3}
            (-1)^{(h-1)/2} [h],\cr
           p_3(z,x)&=zx+x,\cr
           [h]&=T_{h-1}(\sqrt{1+x}),\cr}\e$$
 where $[h]$ expresses the basis element $[h]$ as a polynomial in $x$,
and is used to express $p_2$ as a polynomial in $x$ and $z$. In other words,
$${\cal R}\approx {P[x,z]\over (p_1,p_2,p_3)},\e$$
where $\cal R$ denotes the graph algebra, $P[x,z]$ is the algebra of
polynomials
in $x$ and $z$, and $(p_1,p_2,p_3)$ is the ideal in it generated by
the three polynomials $p_1,p_2$ and $p_3$.

It remains to be seen, now, if this graph algebra satisfies any of
the properties of a fusion ring of an RCFT, in accordance with the
conjecture raised in the introduction. Quite evidently the answer is no!
There are a number of problems. 1) Some of the
structure constants are negative integers. 2) There is no appropriate
symmetric bilinear form. 3) In a related way, the $S$ matrix cannot be made
symmetric. To be more precise let $p_h^j$ be the alleged symmetric bi--linear
form, where since the matrix $p$ has a unique $1$ in each row and column,
it is actually a permutation, expressing a map between exponents $h$, denoted
by $k(j)$, where $j$ is a Dynkin node, and $k(j)$ is an exponent,
and the nodes of the Dynkin diagram. From the properties of RCFT,
the modular matrix must be symmetric, when $p$ is used to lower the index.
Thus, RCFT requires that,
$$v_j^{k(l)}=v_l^{k(j)}.\e$$
More generally, we could allow for a change of normalizations of the
eigenvectors, which are defined only up to a phase, in which case, eq. (30)
assumes the form,
$$\sum_h v_j^h p_{hl}=\sum_h v_l^h p_{hj}.\e$$
It is readily seen that eq. (31), has no solutions for $p_{hj}$,
when $v_j^h$ is taken to be the eigenmatrix of $D_n$, eqs. (25-26). Thus,
this graph algebra is not the fusion ring of any RCFT.

This appears to be quite a catastrophe for the original conjecture we raised,
and in fact, a counter example for it. The question is if there is a way in
which the conjecture we raised could be relaxed, and that this graph
algebra can be
related to an RCFT?  More precisely, we assumed
in our entire discussion, that
the nodes of the graph are the primary fields of the RCFT. There is actually
no reason for this assumption, as we can well build a fusion IRF model
based on non--primary fields \r\Found. Clearly, this entails a change of
basis for the graph algebra. Thus, the question becomes whether the graph
algebra is isomorphic to a fusion ring of an RCFT. Unfortunately, this is
a rather meaningless question, since it is well known that two finite
dimensional algebras are isomorphic if they are both semi--simple, and
have the same dimension. Thus, any graph algebra is isomorphic to any
fusion algebra, provided they have the same number of nodes, respectively,
primary fields. Clearly, this is too weak a criteria to be of
much use.

We conclude the discussion of the $D_n$ cases by noting one particular
basis in which things look rather close to an RCFT. We form the basis,
$$\eqalign{\alpha_h^+&=\half ([h]+[2n-2-h]),\cr
           \alpha_h^-&=(-1)^{(h-1)/2}\half  ([h]-[2n-2-h]),\cr
           z&,\cr }\e$$
where $h=1,3,\ldots,n-3$.
In this basis, the algebra decouples to a direct sum of two subalgebras.
These are the subalgebras generated by $\alpha^+_h$ (along with z, for
even n), and $\alpha^-_h$ (along with $z$, for odd $n$). Denote these two
subalgebras by $A$ and $B$. Then $AB=0$, and $G\approx A\oplus B$.
The question now is any of $A$ and $B$ are the fusion rings of an RCFT?
The fusion rules in this basis become,
$$\eqalign{
\alpha^\pm_r \alpha^\pm_t&=\sum_{s=|r-t|+1\atop s=1\mod 2}^{r+t-1}
\alpha^\pm_s,\cr
z \alpha^+_h&=\cases{0&{\rm odd\ } $n$,\cr
                     z&{\rm even\ } $n$,\cr}\cr
z \alpha^-_h&=\cases{z&{\rm odd\ } $n$,\cr
                     0&{\rm even\ } $n$,\cr}\cr
z^2&=\cases{\displaystyle{\sum_{h=1\atop h=1\mod 2}^{n-2} \alpha_h^+}
&{\rm odd\ }$n$,\cr
\displaystyle{\sum_{h=1\atop h=1\mod 2}^{n-2} \alpha_h^-}&{\rm even\ }$n$.\cr}
\cr}\e$$
It can now be seen that the subalgebra $A$ (generated by $\alpha^+_h$,
for odd $n$, and by $\alpha^-_h$, for even $n$), gives rise to a symmetric
$S$ matrix. Namely, the eigenvectors matrix of this algebra, may be written
as,
$$S_{h,t}=C\sin\left( {\pi ht\over 2n-2}\right),\e$$
for $h,t=1,3,\ldots n-2$, and where $C$ is some constant.
Clearly $S$ is a symmetric unitary matrix, and
the question remains whether it corresponds to an RCFT. Note, that
if we take $n$ to be half integral, this is exactly the $S$ matrix
of the RCFT $SU(2)_{2n-4}/SU(2)_{1/(2n-4)}$. Unfortunately, for an integral
$n$, it can be seen, except for the trivial, $n=4$, not to correspond to an
RCFT, as the equation $(ST)^3=1$ has no solutions with a diagonal matrix
$T$. As this is a necessary condition for $S$ to be a modular matrix,
we conclude that the above $S$ is not the modular matrix of any RCFT.
For $n=5,7$, the entire eigenvalue matrix is symmetric, but still does not
appear to stem from an RCFT. Other basis in which the $S$ matrix is
symmetric can be found.
The significance of these observations remains to be studied. At this
point we conclude that the relation of the $D_n$ models with RCFT is
moot, though short of a counter example to our conjecture.
\par
Let us turn now to the case of the exceptional algebras $E_n$, for
$n=6,7,8$. The exponents of the respective algebras are listed in Table. 1.
The eigenvalues are thus, $\lambda_h=2\cos(\pi h/g)$, where $h$ is any
of the exponents. The eigenvectors are found to be,
$$\cases{v_j=\sin({\pi j h\over g})&for $j\leq n-3$,\cr
         v_n={\sin({3\pi h\over g})\over 2\cos({\pi h\over g})},&\cr
         v_{n-2}=\sin({\pi (n-2) h\over g})-{\sin({\pi (n-3)h\over g})
\over 2\cos({\pi h \over g})},&\cr
         v_{n-1}={v_{n-2}\over 2\cos({\pi h\over g})},&\cr}\e$$
The eigenvectors are further normalized to have absolute value one,
$v_j^h\rarrow$ $ v_j^h/\sqrt{\sum_j (v_j^h)^2}$.
\par
Consider now the case of $E_6$. It is more convenient to define here
the transposed graph algebra associated to the nodes, eq. (11).
We find that all the
structure constants are positive integers. Denoting by $[j]$ the element
of the algebra associated to the $j$th node, we find that the transpose ring
is given by $R\equiv {P[x,y]\over (y^3-2y,x^2-xy-1)}$, where the
Dynkin nodes basis elements of the algebra are given by,
$$[1]=1,\quad,[2]=x,\quad,[3]=xy,\quad,[4]=x(y^2-1),\quad [5]=y^2-1,
\quad [6]=y.\e$$
All the products may be computed from the two relations, $y^3-2y=x^2-xy-1=0$.
The first question now is whether the above graph algebra is a fusion ring.
It is easy to see that this is not the case, and that the matrix of
eigenvectors $v_j^h$ cannot be symmetrized. Thus, although close to the
notion of a fusion IRF model, the $E_6$ IRF model does not stem from an
RCFT.

As in some of the $D_n$ cases we can form combinations of the Dynkin nodes
that give rise to a symmetric matrix of eigenvalues. These are the
combinations $[1]\pm [5]$, $[2]\pm [4]$, $[3]$ and $[6]$ (up to
normalizations). The combination $[1]-[5]$, $[2]-[4]$, generates a subalgebra
which decouples from the rest of the algebra, i.e., the fusion algebra
is the direct sum of the two algebras, $A$ generated by $[1]-[5]$ and
$[2]-[4]$, and the algebra $B$ generated by the rest, $AB=0$. The
corresponding matrix of eigenvalues is symmetric, and thus a candidate for
a modular matrix of an RCFT. The fusion ring $A$ is seen to be isomorphic
to that of $SU(2)_1$ and thus is coming from an RCFT, $R\equiv P[x]/(x^2-1)$.
However, the structure constants of $B$ cannot be made integral, and thus
it cannot be the fusion ring of any RCFT. Again, the significance of
this observation is unclear.

In the case of $E_7$ we find the transpose graph algebra which is,
$$R\equiv {P[x]\over (x^7-6x^5+9x^3-3x)},\e$$
where the elements associated to the Dynkin nodes are given by,
$$\eqalign{
[1]&=1,\cr
[2]&=x,\cr
[3]&=x^2-1,\cr
[4]&=x^3-2 x,\cr
[5]&=x^6-5 x^4+5 x^2,\cr
[6]&=x^5-5 x^3+5 x,\cr
[7]&=-x^6+6 x^4-8 x^2+1.\cr}\e$$
Again, all the structure constants are non--negative integers, in terms
of the Dynkin basis elements, $M_{ij}^k\geq0$.

In the case of $E_8$, we find the the transposed graph ring again has all the
structure constants as positive integers. The ring is given by
$$R\equiv {P[x]\over (
1 - 8 x^2 + 14 x^4 - 7 x^6 + x^8)}.\e$$
The Dynkin elements are now given by,
$$\eqalign{
[1]&=1,\cr
[2]&=x,\cr
[3]&=-1 + x^2,\cr
[4]&=-2 x + x^3,\cr
[5]&=1 - 3 x^2 + x^4,\cr
[6]&=-2 x + 9 x^3 - 6 x^5 + x^7,\cr
[7]&=-2 + 9 x^2 - 6 x^4 + x^6,\cr
[8]&=5 x - 13 x^3 + 7 x^5 - x^7,\cr}\e$$
The polynomials we find for $E_n$, $n=6,7,8$, may be considered as the
generalizations of the Chebishev polynomials, which arise for $A_n$, to
the exceptional algebras.

A very interesting property of all the $E_n$ graph algebras we find is
that the product with $x=[2]$ gives back the incidence matrix of the
graph,
$$x\times[n]=\sum_m M_{nm} [m],\e$$
where $M_{nm}$ is the incidence matrix of the graph, which is the
Dynkin diagram of the respective algebra. Further, this rule alone,
determines uniquely the entire algebra. Thus, the rings we found
are exactly those giving the Dynkin graph as an admissibility
condition. Namely, the $E_n$ models can be thought of as
the models IRF$(R,x)$, where $R$ stands for the graph ring, and
$x=[2]$ is the element used in the admissibility condition.
The fact that we managed to lift the admissibility relation to a
full `fusion ring' with positive integer structure constants is
non--trivial, and may be connected with the solvability of the
model.

In all the $E_n$ cases, the matrix of eigenvalues is non--symmetric, and
thus does not correspond directly to an actual RCFT. As noted earlier,
they are certainly isomorphic to fusion rings of RCFT, but this is a
somewhat meaningless fact. As in the $D_n$ case, we can also form the
graph algebra, based on the exponents, eq. (10). Forming as in the $D_n$ case,
the combinations, $A_-=\half ([h]-[g-h])$ and $A_+=\half([h]+[g-h])$,
where $h$ is any of the exponents,
we find
that the two subalgebras decouple, $A_-A_+=0$. For $E_7$ this leads also to a
symmetric eigenvalue matrix, and fusion rules which are integers,
and thus are full candidates for an RCFT. It remains to explore further whether
an RCFT based on this can be built. For $E_6$ and $E_8$ we find non--symmetric
eigenvalue matrix, indicating that the two subalgebras do not represent an
RCFT.

In conclusion, we studied here a variety of examples of solvable IRF models,
to judge if they stem from an RCFT. We formed graph algebras
based on the
eigenvalues of the admissibility conditions. If a theory stems from an
RCFT its graph algebra must be the fusion rules of the model. This also
gives immediately the solutions to the STE, eq. (2), by an extension
of the braiding matrices of the theory, as described in ref. \r\Found.
We studied here two families of examples, the Gordon generalized (GG), KAW
and grand hierarchies,
and the Pasquier $D$--$E$ models.
Remarkably, it was shown that all the hierarchy models models stem from
some RCFT
related to $SU(2)$ current algebra. On the other hand, the $D-E$ models were
seen not to correspond directly to an RCFT. It remains to be studies, whether
these models stem from a mixture of primary fields in an RCFT. While, not
ruling the possibility out, we found that, in general, there does not seem
to be a natural way to relate these models to an RCFT. The most likely
conclusion to be drawn is that while most solvable IRF models studied to date
come from some RCFT, other solutions to the STE exist, which do not appear
to be related to an RCFT in the manner described in ref. \r\Found, with
the $D$ and $E$ models as examples. The question certainly requires further
study.

We hope that we have further illuminated here the connection between RCFT
and solvable lattice models. A great host of models, erratically constructed
previously, all stem from the unified construction described in ref. \r\Found.
While few exceptions were found, it remains to study how these can be fitted
into the general framework.
\ACK
I thank P. Di--Francesco, W. lerche and J.B. Zuber for helpful
comments. While writing this work, I received
\REF\Zuber{P. Di Francesco, F. Lesage and J.B. Zuber, Saclay preprint
SPhT 93/057, June (1993)}\r\Zuber, which somewhat relates to the present
work.
\refout
\bye